\documentclass[11pt]{article}
\input epsf
\usepackage{moriond,epsfig}

\def\Journal#1#2#3#4{{#1} {\bf #2}, #3 (#4)}

\def\PLB{{\em Phys. Lett.}  B}

\def\PRD{{\em Phys. Rev.} D}

\def\EPJ{{\em Eur. Phys. J.} {\bf C}}
\def\PREP{\em Phys. Rept.}

\def\RNC{\em Riv. Nuovo Cim.}

\newcommand{\gevcc}{\ensuremath{{\rm GeV}\!/c^2}}

\newcommand{\epemto}{\ensuremath{{{\rm e}^+{\rm e}^- \to}}}

\newcommand{\st}{\ensuremath{{\tilde{\rm t}}}}

\newcommand{\sq}{\ensuremath{{\tilde{\rm q}}}}
\newcommand{\glu}{\ensuremath{{\tilde{\rm g}}}}
\newcommand{\mst}{\ensuremath{m_\st}}

\newcommand{\mglu}{\ensuremath{m_\glu}}

\newcommand{\sqsqbar}{\ensuremath{\sq\bar{\sq}}}
\newcommand{\qqbar}{\ensuremath{{\rm q\bar{q}}}}

\def\be{\begin{equation}}
\def\ee{\end{equation}}
\def\bea{\begin{eqnarray}}
\def\eea{\end{eqnarray}}

\begin{document}
\vspace*{4cm}
\title{Stop and sbottom searches at LEP}

\author{ A.C. Kraan }

\address{Niels Bohr Institute, Blegdamsvej 17, 2100 K\o benhavn-\o,\\
Denmark}

\maketitle\abstracts{
This talk reviews the searches for sbottom and stop quarks at LEP. The existing results of the four LEP experiments for sbottom and stop quarks searches are summarized. Furthermore, new mass limits on stable hadronizing squarks and gluinos are derived by combining Z-lineshape constraints with direct searches for squarks and gluinos using ALEPH data. All limits are derived in the framework of R-parity conserving models.}

\section{Introduction}
Supersymmetry (SUSY) is one of the leading candidates for theories predicting physics beyond the standard model (SM). In supersymmetric theories \cite{susy}, every particle has a supersymmetric 'partner', differing in spin by half a unit. Since no SUSY particles have been observed so far, we know that, if SUSY exists, the symmetry is broken, and the masses of the SUSY particles are much larger than their SM partners, probably beyond the reach of LEP. However a special situation arises for the third generation squarks (the SUSY partners of quarks), for which substantial mixing occurs between the ``left-handed'' and ``right-handed'' partners. This results in a heavy and a light squark mass eigenstate, the latter possibly in the reach of LEP. 

In R-parity conserving supersymmetry models, the lightest supersymmetric particle (LSP) cannot decay. The phenomenology crucially depends on this LSP. In 'standard' SUSY scenarios, in which the electric, weak and strong couplings unify at the GUT scale, the LSP is either a neutralino (a mixture of the superpartners of the gauge bosons) or sneutrino (the superpartner of the neutrino), while the gluino (the superpartner of the gluon) is naturally heavy. However, there are SUSY scenarios in which the gluino or the squark is LSP, in which case the LSP hadronizes into stable particles called R-hadrons.

In this presentation, the most important results are reviewed from searches for stops and sbottom quarks at LEP. For LEP\,1, this corresponds to 4.5 million Z-decays per experiment, while during LEP\,2 circa 775 pb$^{-1}$ of data per experiment was collected. 

First, stop and sbottom searches in standard SUSY scenarios are summarized. When available, the SUSY LEP working group results are used, which are based on combinations of ALEPH, DELPHI, L3 and OPAL (ADLO). This is followed by stop and sbottom searches with a gluino LSP, done by DELPHI. After that, new limits on gluinos and squarks are derived by combining Z-lineshape constraints with direct searches at LEP\,1 and LEP\,2 with ALEPH data.  
\section{Stop and sbottom searches: status 2002}
Searches for stops and sbottoms decaying into a quark and a non-interacting LSP (neutralino or sneutrino), have been done by the four LEP experiments \cite{squarks}. In all decay channels, searches are based on a large missing energy signal due to the escaping LSP. The missing energy is closely related to the mass difference $\Delta$M between the squark and the LSP: small $\Delta$M values result in a large missing energy signal and vice versa. It must be noted that this also applies in the case in which the gluino is LSP. Although in that case some hadronic interaction occurs in the calorimeters, the missing energy is still substantial due to the missing mass of the particle and the poor hadronic interaction in the detector, provided that the gluino is heavy enough. In the decay channels $\tilde{t}\rightarrow c \tilde{\chi}^o$,  $\tilde{t}\rightarrow c \tilde{g}$, $\tilde{b}\rightarrow b + \tilde{\chi}^o$, and $\tilde{b}\rightarrow b + \tilde{g}$, the search topology is missing energy and acoplanar quark jets. In the channel $\tilde{t}\rightarrow b\l\nu$, isolated leptons are additionally searched for. In the four-body decay channel $\tilde{t}\rightarrow b\bar{f}_1f_2\tilde{\chi}^o$, studied by ALEPH, the topology is multi-jets and missing energy. No excess of events was observed in any channel. In figure 1 excluded regions are shown in the (squark mass - LSP mass) plane. Only squark masses below 50 \gevcc\ are displayed, since lower values have been excluded at LEP\,1. The case, in which the mass differences $\Delta$M between the squark and the LSP is so small that the squark has a sizable lifetime, is only addressed by ALEPH \cite{Barate:2000qf}, and resulted in an absolute limit on the stop mass of 63 \gevcc, if the LSP is sneutrino or neutralino. Figure 1 also display the results for the decay channels $\tilde{t}\rightarrow c + \tilde{g}$ and  $\tilde{b}\rightarrow b + \tilde{g}$, which until 2002 had only been studied by DELPHI \cite{delphi}. However, even when combining with other searches for stable gluinos \cite{gunion}, there are still regions in parameter space open in this channel: the small gluino mass region, the region with squark masses below 50 \gevcc (the LEP\,1 limit does not apply since this limit is based on a non-interacting LSP) and the region in which the mass difference between gluino and squark is smaller than 5 \gevcc. In the next section we will describe how to cover these regions. 
\begin{figure}[t]
\epsfig{file=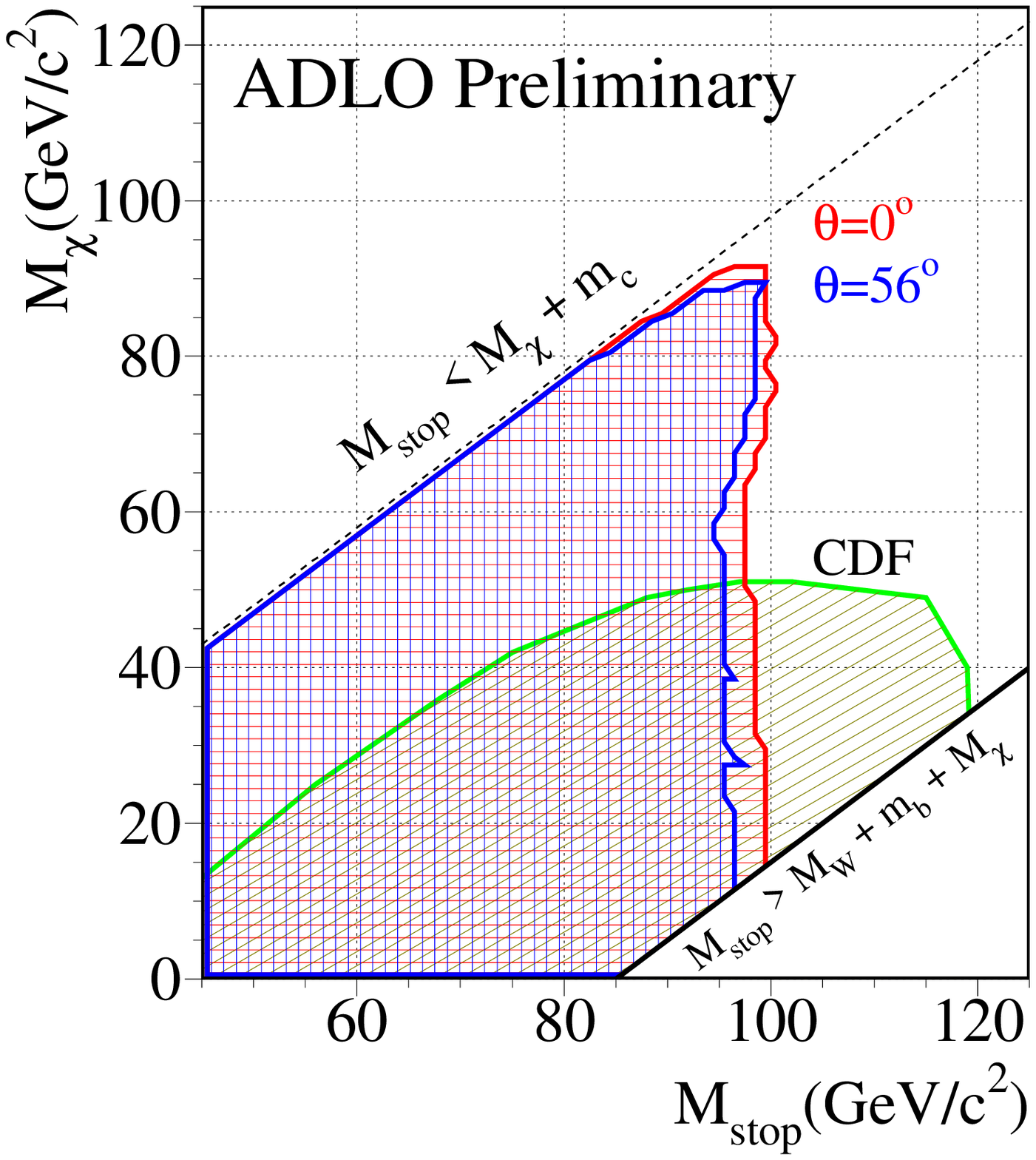,height=6cm,width=6cm}\hspace*{-0.7cm} \epsfig{file=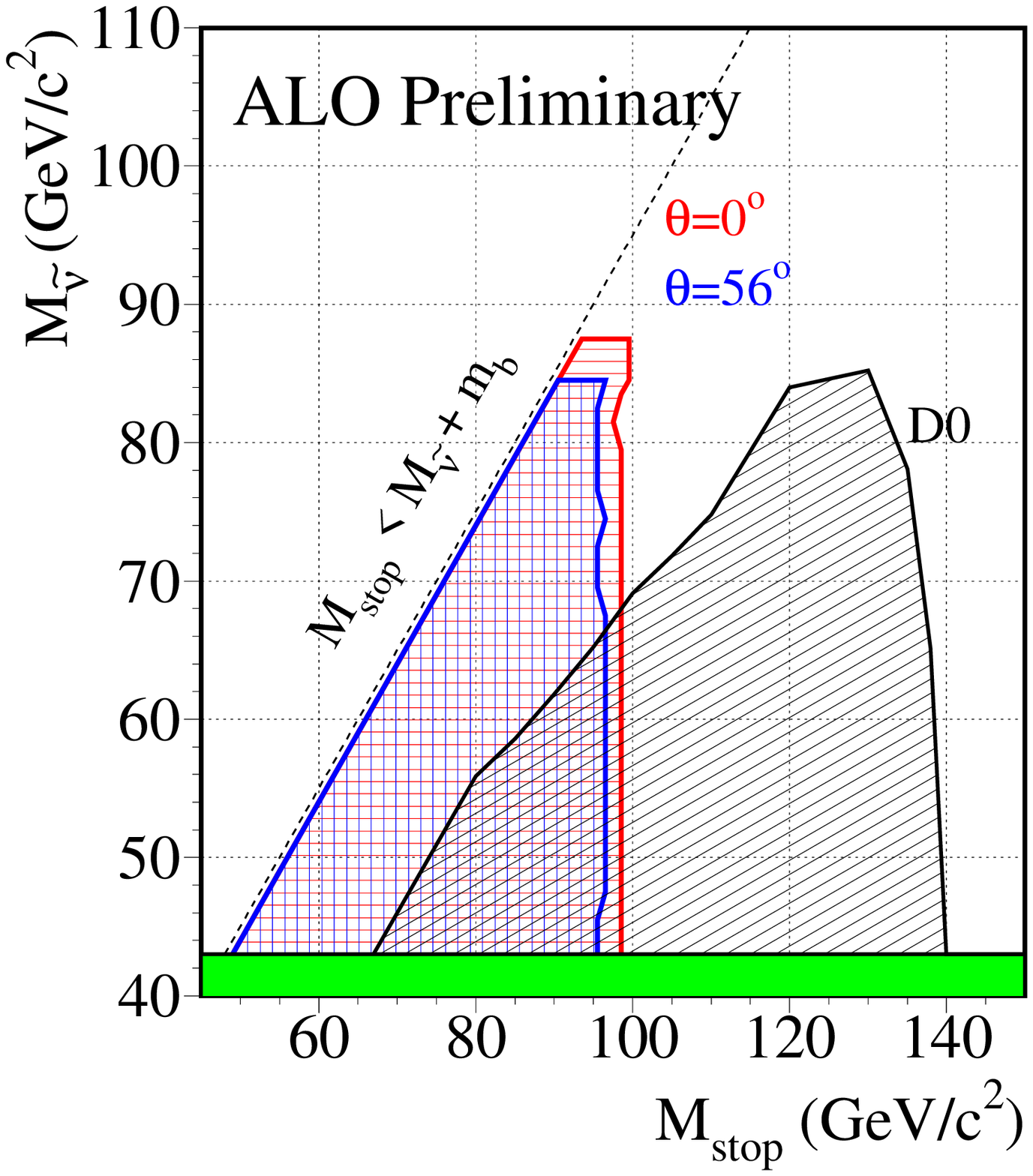,height=6cm,width=6cm}\hspace*{-0.7cm}\epsfig{file=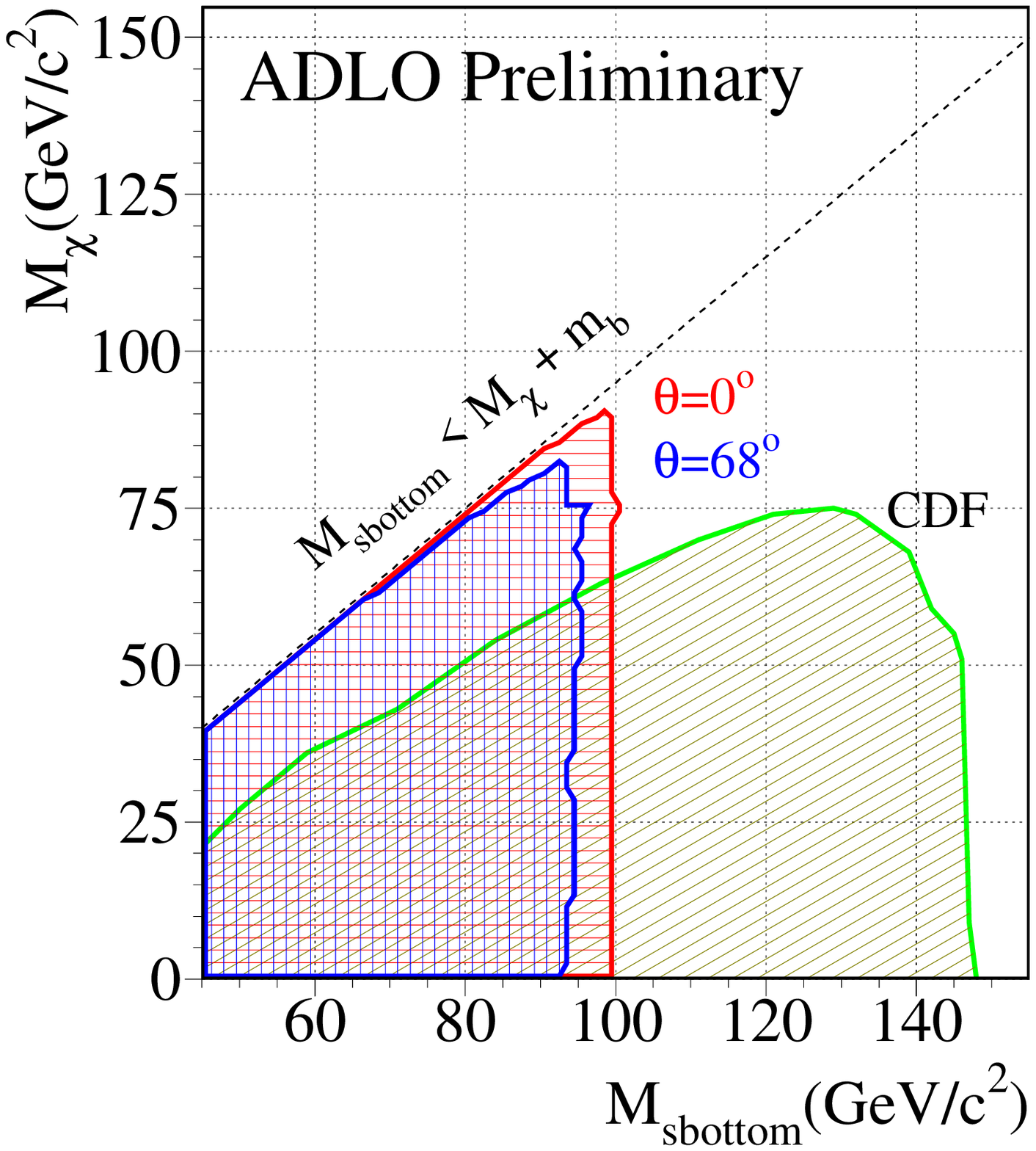,height=6cm,width=6cm}

 \hspace*{0.2cm}\epsfig{file=giacomo.epsi,height=4.9cm,width=4.9cm} \hspace*{0.4cm}\epsfig{file=peq05.epsi,height=4.8cm,width=4.8cm} \hspace*{0.4cm}\epsfig{file=peq05sbot.epsi,height=4.8cm,width=4.8cm}\label{fig:excl}
\caption{\footnotesize The three top figures show the excluded regions for the decay channels $\tilde{t}\rightarrow c \chi_O$ (top left),  $\tilde{t}\rightarrow b\l \tilde{\nu}$ (top center) and  $\tilde{b}\rightarrow b \tilde{\chi}$ (top right), as obtained by the SUSY LEP working group. The three bottom figures show relevant results as obtained by ALEPH and DELPHI: $\tilde{t}\rightarrow b\bar{f}_1f_2\tilde{\chi}^o$ (bottom left), $\tilde{b}\rightarrow b + \tilde{\chi}^o$ (bottom center) and  $\tilde{t}\rightarrow c \tilde{g}$ (bottom right). All excluded regions (the shaded areas) are at 95\%\,C.L.}
\end{figure}

\section{New results for squarks and gluinos}
\subsection{Z-lineshape constraints}\label{sec:z}
Constraints on new physics can be set using Z-lineshape constrains. For example, when the gluino mass is small, the \epemto\ \qqbar\glu\glu\ production 
cross section is large enough to sizable contribute to the Z hadronic 
width. The accurate electroweak measurements at LEP and SLC allow a 
model-independent upper limit of 3.9\,MeV to be set on the Z width 
for purely hadronic final states. All gluino masses below 
6.3\,\gevcc, irrespective of the gluino decay and hadronization mechanisms,
are therefore excluded at 95\% confidence level \cite{janot}. This defenitely closed the long discussed 'light gluino mass window'. In the same way, the production cross section for  \epemto\ \qqbar\glu\glu\  allows to exclude all squark masses below 1.3\,\gevcc\ at 95\% confidence level. 
\subsection{LEP\,1 searches for stable gluinos and stable squarks}\label{sec:1}
At LEP\,1, squarks and gluinos can be produced via gluon splitting. For larger masses than the masses excluded above, a combination of previously developed selections \cite{lep1,dudu} applies, again based on acoplanar jets and missing energy. In combination with the Z-lineshape constraints, this results in a 95\% confidence level exclusion of squark masses below 15.7\,\gevcc\ and gluino masses below 26.9\,\gevcc.
\subsection{LEP\,2 search for stable squarks}\label{sec:2}
Since squarks decouple from the Z at certain mixing angles, LEP\,2 data have been analyzed to search for stable squark. Stable stops hadronize into stop R-hadrons. Issues which are addressed in the Monte Carlo simulation of the signal are stop hadronization and the R-hadronic interaction in the detector. The usual search for heavy stable particles applies \cite{Barate:2000qf}. In combination with the LEP\,1 search for squarks and the Z-lineshape constraints, the search allows stable stop masses below 95\,\gevcc\ and stable sbottom masses between below 92\,\gevcc\ to be excluded at 95\%\,C.L.
\subsection{LEP\,2 search for stops decaying into stable gluinos}\label{sec:3}
If the gluino is the LSP and the lighter scalar top quark the NLSP, the 
stop decays into a gluino and a c or a u quark. The decay 
widths~\cite{Hikasa:1987db} are such that the stop has time to hadronize before it decays. The signal Monte Carlo simulation addresses stop hadronization, stop decay inside the hadron, gluino hadronization into R-hadrons, and the R-hadronic interaction in the detector. The search topology is again missing energy and acoplanar jets. In addition, the case in which the stop has a sizable lifetime, relevant when $\Delta M \simeq m_{\rm D}$,  is treated in the same way as described elsewhere \cite{Barate:2000qf}. 


\begin{figure}[t]
\begin{center}
\epsfig{file=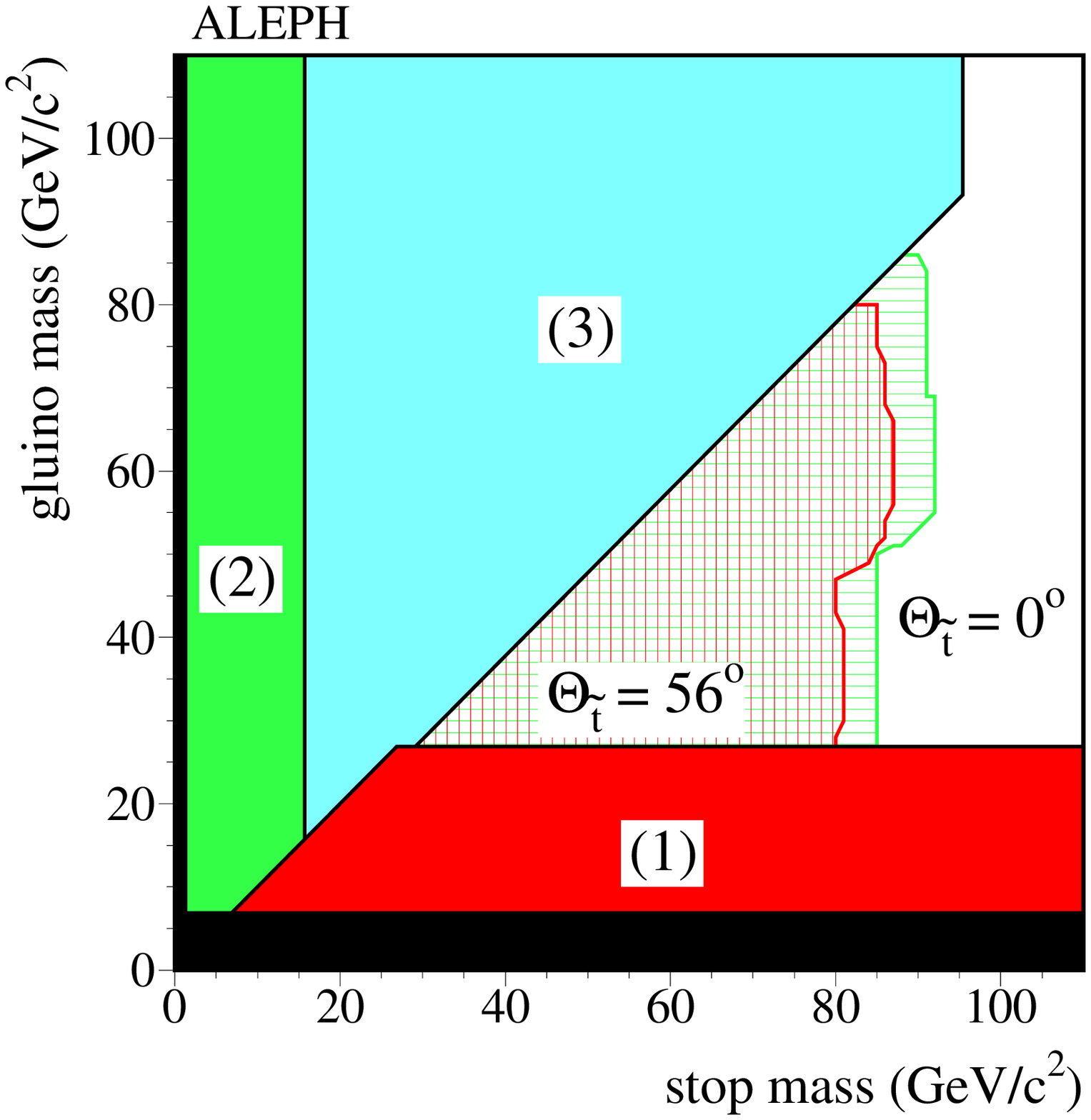,height=7cm,width=7cm}
\end{center}
\caption{\footnotesize  The 95\%\,C.L. excluded regions in the plane 
(\mst, \mglu) from the combination of all searches. The black area at very small gluino and squark masses is excluded by the precise measurement of the Z-lineshape; Regions (1), (2) and (3) are excluded by the search for \epemto\ \qqbar\glu\glu\ at LEP\,1, for \epemto\ \qqbar\sqsqbar\ at LEP\,1, and for heavy stable charged particles from \sqsqbar\ production at LEP\,2, respectively. The hatched areas are  excluded by the acoplanar jet plus missing energy search at LEP\,2. \label{fig:all}}
\end{figure}
\section{Conclusion}
Squarks have been extensively searched for at LEP in the framework of R-parity conserving models with neutralino, sneutrino and gluino LSP. No excess of events is observed in any channel. 
 
\section*{Acknowledgments}
I would like to thank Patrick Janot, Giacomo Sguazzoni and Mario Antonelli for their contributions to the analysis and for their constant help and support. Furthermore I thank Torbj\"orn Sj\"ostrand for his support with the Monte Carlo simulation and for valuable discussions and advice.

\end{document}